\documentclass[10pt,a4paper]{article}
\makeatletter
\AtBeginDocument{%
  \@ifpackageloaded{hyperref}
  {\def\@doi#1{\href{https://doi.org/#1}
      {\ttfamily https://doi.org/#1}\egroup}}
  {\def\@doi#1{\ttfamily https://doi.org/#1\egroup}}
  \def\doi{\bgroup\catcode`\_=12\relax\@doi}}
\makeatother
\usepackage{authblk}
\usepackage{amsfonts}
\usepackage{amsmath}
\usepackage{amssymb}
\usepackage[english]{babel}
\usepackage[justification=centering]{caption}
\usepackage[T1]{fontenc}
\usepackage{graphicx}
\usepackage{ifsym}
\usepackage[utf8]{inputenc}
\usepackage{marvosym}
\usepackage{stmaryrd}
\usepackage{subfigure}
\usepackage{hyperref}
\usepackage{graphicx}
\usepackage{xcolor}
\usepackage{mathrsfs}
\usepackage{wrapfig}
\usepackage{mathpartir}
\usepackage{pgfplotstable}
\usepackage{booktabs} 
\usepackage{calculator}
\usepackage[binary-units=true]{siunitx}

\hypersetup{
  colorlinks   = true, 
  urlcolor     = blue, 
  linkcolor    = blue, 
  citecolor   = red 
}
\usepackage[ruled,linesnumbered]{algorithm2e}

\SetKwProg{Fn}{function}{}{}
\SetKwFor{Forever}{forever}{}{}
\DontPrintSemicolon

\definecolor{blue}{RGB}{0, 102, 204}
\definecolor{brick-red}{RGB}{203, 65, 84}
\definecolor{brown}{RGB}{216, 203, 175}
\definecolor{green}{RGB}{0, 153, 76}
\definecolor{grey}{RGB}{151, 100, 100}
\definecolor{orange}{RGB}{255, 166, 48}
\definecolor{purple}{RGB}{102, 0, 204}
\definecolor{red}{RGB}{202, 3, 3}

\usepackage{tikz}
\usetikzlibrary{automata}
\usetikzlibrary{arrows}
\usetikzlibrary{calc}
\usetikzlibrary{backgrounds}
\usetikzlibrary{chains}
\usetikzlibrary{decorations.pathreplacing}
\usetikzlibrary{decorations.text}
\usetikzlibrary{fadings}
\usetikzlibrary{shapes}

\tikzstyle{cube}=[
   thick, fill=brown, regular polygon, regular polygon sides=3, rounded
   corners=1pt, inner sep=0mm, outer sep=0mm, minimum size=1cm
]
\tikzstyle{gp}=[-,thick,color=black]
\tikzstyle{overcube}=[cross out,draw=red,solid,ultra thick]
\tikzstyle{move}=[
   ->,dotted, >=latex, rounded corners=5pt, thick,draw=blue, line width=0.5mm
]

\renewcommand{\bf}[1]{\textbf{#1}}

\renewcommand{\tt}[1]{\texttt{#1}}

\newcommand{\mcc}[0]{\tt{mcc}}

\newcommand{\vv}[1]{\tt{#1}}

\newcommand{\opp}[0]{\texttt{++}}
\newcommand{\omm}[0]{\texttt{-\kern -0.1em-}}

\usepackage{array}
\usepackage{booktabs}

\newcolumntype{L}[1]{>{\raggedright\let\newline\\\arraybackslash\hspace{0pt}}m{#1}}
\newcolumntype{C}[1]{>{\centering\let\newline\\\arraybackslash\hspace{0pt}}m{#1}}
\newcolumntype{R}[1]{>{\raggedleft\let\newline\\\arraybackslash\hspace{0pt}}m{#1}}

\providecommand{\keywords}[1]
{
  \small	
  \textbf{\textit{Keywords---}} #1
}
\title{MCC: a Tool for Unfolding Colored\\ Petri Nets in PNML Format}
\author[1]{Silvano {Dal Zilio}}
\affil[1]{LAAS-CNRS, Universit\'{e} de Toulouse, CNRS, Toulouse, France}
\date{}
\begin{document}

\maketitle
\sloppy

\begin{abstract}
  MCC is a tool designed for a very specific task: to transform the
  models of High-Level Petri nets, given in the PNML syntax, into
  equivalent Place/Transition nets. The name of the tool derives from
  the annual Model-Checking Contest, a competition of model-checking
  tools that provides a large and diverse collection of PNML
  models. This choice in naming serves to underline the main focus of
  the tool, which is to provide an open and efficient solution that
  lowers the access cost for developers wanting to engage in this
  competition.
  
  We describe the architecture and functionalities of our tool and
  show how it compares with other existing solutions. Despite the fact
  that the problem we target is abundantly covered in the literature,
  we show that it is still possible to innovate. To substantiate this
  assertion, we put a particular emphasis on two distinctive features
  of MCC that have proved useful when dealing with some of the most
  challenging colored models in the contest, namely the use of a
  restricted notion of ``higher-order
  invariant'', and the support of a Petri net scripting language.\\

  \noindent\keywords{Tools, PNML, High-Level Petri nets, Colored Petri
    nets}
\end{abstract}

\section{Introduction}

The Petri Net Markup Language
(PNML)~\cite{billington2003petri} is an
XML-based interchange format for representing Petri nets and their
extensions. One of its main goal is to provide developers of Petri net
tools with a convenient, open and standardized format to exchange and
store models. While its focus is on openness and extensibility, the
PNML spotlights two main categories of models: standard
Place/Transition nets (P/T nets), and a class of Colored Petri nets,
called {High-Level Petri Nets} (HLPN), where all types have finite
domains and expressions are limited to a restricted set of
operators~\cite{chiola1991well,jensen1987coloured}.

In this paper we present \mcc, a tool designed for the single task of
\emph{unfolding} the models of High-Level Petri nets, given in the
PNML syntax, into equivalent Place/Transition nets. The name of the
tool derives from the annual Model-Checking Contest
(MCC)~\cite{mcc2019}, a competition of ``Petri tools'' that
makes an extensive use of PNML and that provides a large and diverse
collection of PNML models, some of which are colored. Our choice when
naming \mcc\ was to underline the main focus of the tool, which is to
provide an open and efficient solution that lowers the access cost for
developers wanting to engage in the MCC.

We seek to follow the open philosophy of PNML by providing a software
that can be easily extended to add new output formats. Until recently,
the tool supported the generation of Petri nets in both the
TINA~\cite{tina} (.net) and LOLA~\cite{lola} formats; but it has been
designed with the goal to easily support new tools. To support this
claim, we have very recently added a new command to print the
resulting P/T net in PNML format. This extension to the code serves as
a guideline for developers that would like to extend \mcc\ for their
need.

The rest of the paper is organized as follows. In
Sect.~\ref{sec:inst-supp-oper}, we describe the basic functionalities
of \mcc\ and give an overview of the PNML elements supported by our
tool, we also propose three new classes of colored models that are
representative of use cases found in the MCC
repository~\cite{HillahK17}. Next, we describe the architecture of
\mcc\ and discuss possible applications of its libraries. Before
concluding, we compare \mcc\ with other existing solutions. Despite
the fact that the problem we target is abundantly covered in the
literature, we show that it is still possible to innovate. We describe
two particular examples of optimizations that have proved useful when
dealing with some of the most challenging colored models in the
contest, namely the use of a restricted notion of ``higher-order
invariant'', and the support of a Petri net scripting language.

\section{Installation, Usage and Supported PNML Elements}
\label{sec:inst-supp-oper}

The source code of \mcc\ is made freely available on
GitHub\footnote{See \url{https://github.com/dalzilio/mcc} for the
  source code. Binaries for Windows, Linux and MacOS are available at
  \url{https://github.com/dalzilio/mcc/releases}} and is released as
open-software under the CECILL-B license; see
\url{https://github.com/dalzilio/mcc}. The code repository also
provides a set of PNML files taken from the open collection of models
from the MCC~\cite{HillahK17}.These files are provided in the source
code repository to be used for benchmarking and continuous
testing. The tool can also be easily compiled, from source, on any
computer that provides a recent distribution of the Go programming
language.

\subsubsection*{Basic usage.}
Tool \mcc\ is a command-line application that accepts three primary
subcommands: \vv{hlnet}, \vv{lola} and \vv{pnml}. In this paper, we
focus on the \vv{mcc hlnet} command, that generates a Petri net file
in the TINA \emph{net} format~\cite{tina}. Similarly, commands
\vv{lola} and \vv{pnml} generate an equivalent output but targeting,
respectively, the LoLa~\cite{lola} and PNML formats for P/T nets.

We follow the UNIX philosophy and provide a small program, tailored
for a precise task, that can be composed using files, pipes and shell
script commands to build more complex solutions.  As it is customary,
option \verb+-h+ prints a usage message listing the parameters and
options accepted by the command.

The typical usage scenario is to provide a path to a PNML file, say
\verb+model.pnml+, and invoke the tool with a command such as
``\verb+mcc hlnet -i model.pnml+''. By default, the result is written
in file \verb+model.net+, unless option \vv{-o} or \vv{-{}-name} is
used. We discuss some of the other options of \mcc\ in the sections
that follow.

\subsubsection*{PNML elements supported by MCC.}
The input format supported by \mcc\ covers most of the PNML syntax
defined in the ISO/IEC 15909-2 standard, which corresponds to the
definition of HLPN.

High-Level Petri nets form a subset of colored nets defined by a
restriction on the types and expressions that are allowed in a
net~\cite{billington2003petri,hillah2006pn}.
The core action language of HLPN is a simple, first-order declarative
language organized into categories for types, values and
expressions. Essentially, HLPN is built around a nominal type system
where possible ground types include a constant for ``plain tokens''
(\vv{dot}), and three different methods for declaring finite, ordered
enumeration types (finite, cyclic, and integer range). The type system
also includes ``product types'', used for tuples of values, and a
notion of \emph{partition elements}, which are (named) subsets of
constants belonging to the same type.

Expressions are built from values and operations and describe
multisets of colors, which act as the marking of places. For instance,
the language include operators \vv{add} and \vv{subtract}, that
correspond to multiset union and difference. The language also
includes a notion of \emph{patterns}, which are expressions that
includes variables (in a linear way), and of \emph{conditions}, which
are boolean expressions derived from a few comparison operators.  A
simple way to describe the subset of the PNML standard supported in
\mcc\ is to list the XML elements supported in each of these
categories (most of the element
names are self-explanatory):\\

\begin{tabular}{lrl}
  types &::=& \vv{dot} $\mid$ \vv{cyclicenumeration} $\mid$
              \vv{finiteenumeration}\\
        &$\mid$& \vv{finiteintrange} $\mid$  \vv{productsort}\\
        &$\mid$& \vv{partition} $\mid$ \vv{partitionelement}\\
  values &::=& \vv{dotconstant} $\mid$ \vv{feconstant} $\mid$
               \vv{finiteintrangeconstant}\\
  expressions &::=&  \vv{variable} $\mid$ \vv{successor}   $\mid$
                    \vv{predecessor}  $\mid$ \vv{tuple}\\
        &$\mid$& \vv{all} 
                 $\mid$ \vv{add} $\mid$ \vv{subtract}\\
  conditions &::=& \vv{or}  $\mid$ \vv{and}  $\mid$
                   \vv{equality} $\mid$ \vv{inequality}\\
        &$\mid$& \vv{lessthan}  $\mid$ 
                 \vv{greaterthan}\\
        &$\mid$& \vv{greaterthanorequal}  $\mid$ \vv{lessthanorequal}\\
\end{tabular}\\

\begin{figure}
  \begin{minipage}[b]{0.275\textwidth}
    \includegraphics[width=\textwidth]{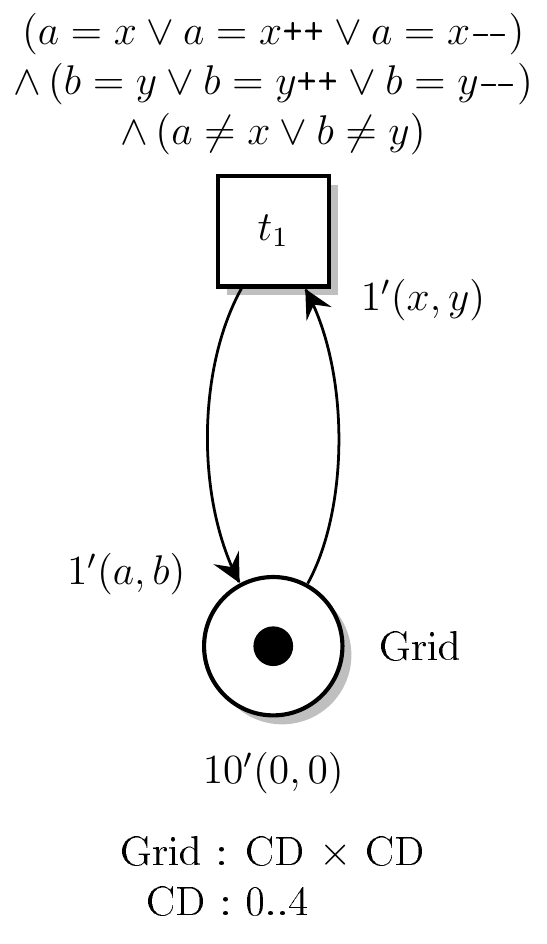}
    \caption{Diffusion}
    \label{fig:1}
  \end{minipage}
  \hfill
  \begin{minipage}[b]{0.39\textwidth}
    \includegraphics[width=\textwidth]{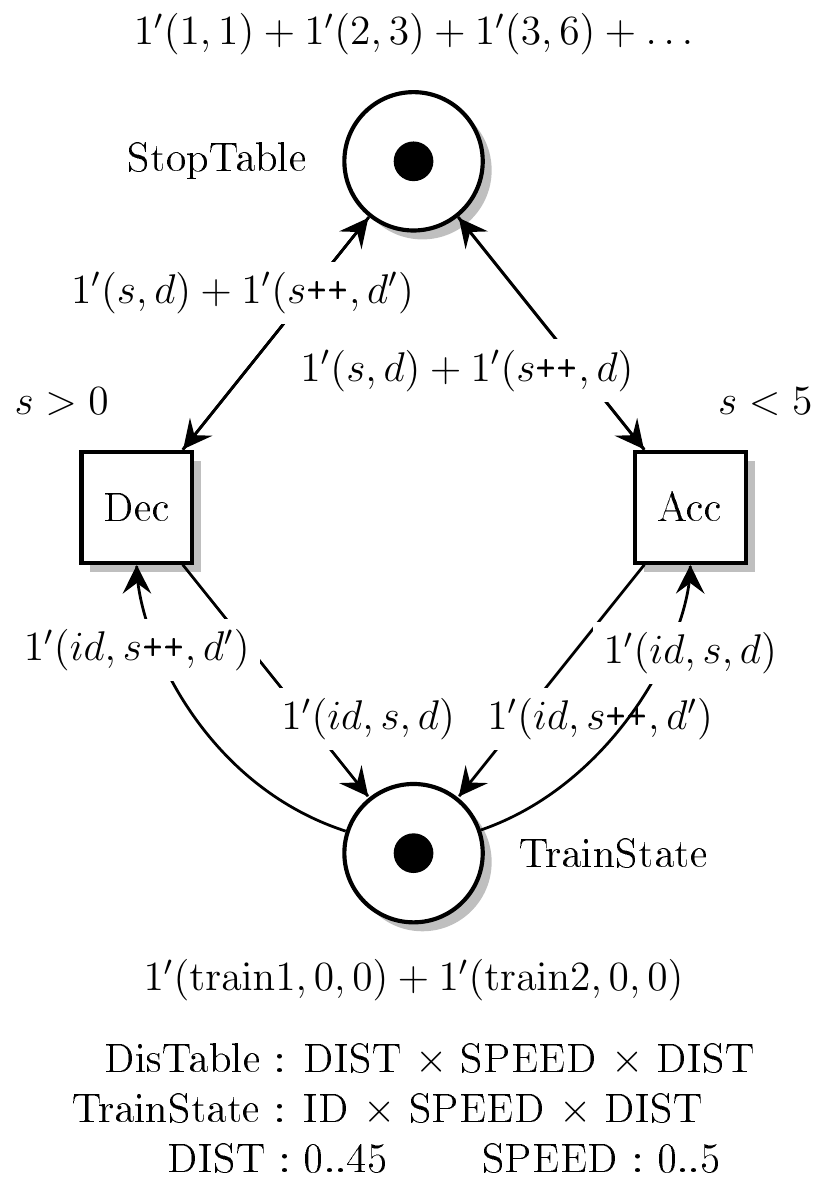}
    \caption{TrainTable}
    \label{fig:2}
  \end{minipage}
  \hfill
  \begin{minipage}[b]{0.3\textwidth}
    \includegraphics[width=\textwidth]{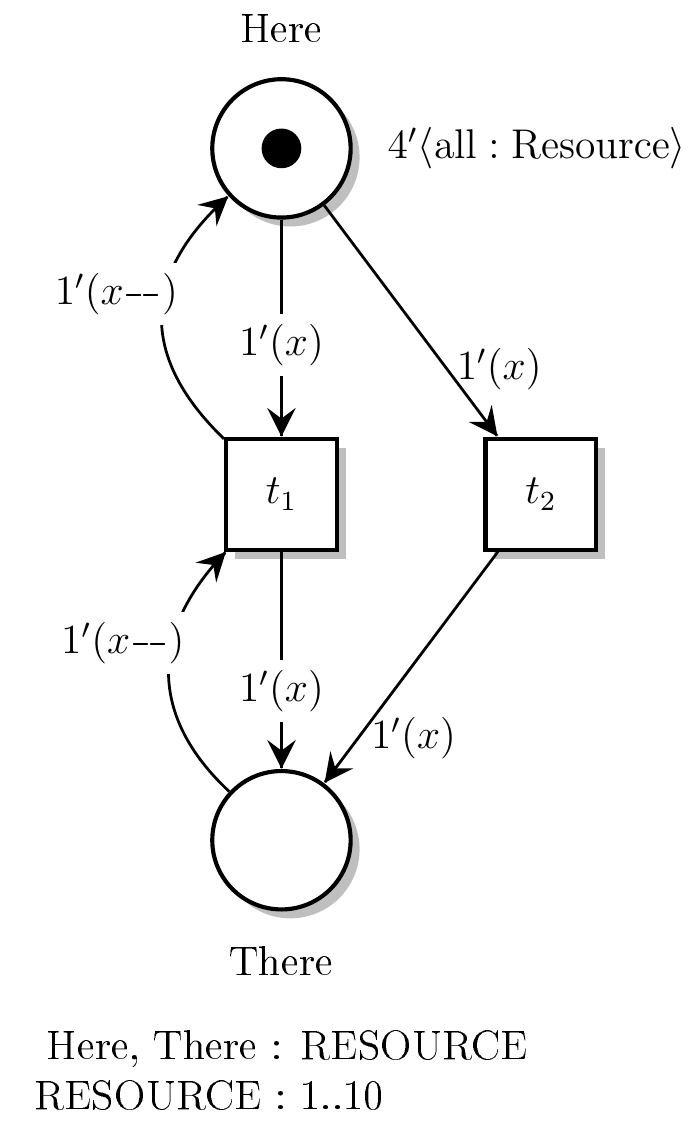}
    \caption{Resource Swap}
    \label{fig:3}
  \end{minipage}
\end{figure}

The \mcc\ tool, in its latest version, supports all the operators used
in models of the Model-Checking Contest. To better understand this
fragment, we give three examples of HLPN that can be expressed using
these constructs, see Fig.~\ref{fig:1} to~\ref{fig:3}. Each of these
examples illustrate an interesting class of parametric models found in
the MCC and will be useful later to discuss the strengths and
weaknesses of our approach.  None of these models are part of the MCC
repository (yet), but their PNML specification can be found in the
\href{https://github.com/dalzilio/mcc/tree/master/docs/models}{\mcc\
  source code repository}.

\subsubsection*{Three representative examples.}
Our first example, Fig.~\ref{fig:1}, illustrates the use of colors to
model a complex network topology. While Diffusion is not part of the
MCC repository, it is the colored equivalent of model {Grid2d}; it is
also the main benchmark in~\cite{liu2012efficient}. In this model,
values in the place {Grid} are of the form $(x, y)$, with
$x, y \in 0..4$. Hence we can interpret colors as cells on a
$5\times5$ grid and values as ``tokens'' in these cells. Tokens can
move to an adjacent cell by firing transition $t_1$ but cannot cross
borders. (In our diagrams we use $\opp$ for \vv{successor}, $>$ for
\vv{greaterthan}, and $+$ for \vv{add}.) All the behavior is
concentrated on the condition associated with $t_1$. Since the
expression contains four variables; we potentially have
$|\mathrm{CD}|^4$ different ways to enable $t_1$.

{TrainTable}, is an example where colors are used to simulate complex
relations between data values. Place {StopTable} is initialized with a
list of pairs associating, to each (integer) speed in $0..5$, the
safety distance needed for a train to stop. Hence TrainTable tabulates
a non-linear constraint between speed and distance. Place {TrainState}
stores the current state of two different trains. Each time a train
accelerate (Acc), or decelerate (Dec), the safety distance is
updated. TrainTable is a simplified version of the BART model. We can
make this model more complex by storing the distance traveled instead
of the safety distance (Traintable-Dist); or even more complex by
storing both values (TrainTable-Stop+Dist).

Our last example, Swap, is typical of systems built from the
composition of multiple copies of the same component and where
interactions are limited to ``neighbors''. The model obeys some
interesting syntactical restrictions: it does not use conditions on
the transitions and inscriptions on arcs are limited to two patterns,
$x$ or ${x}\omm$. This is representative of many models, such as the
celebrated \emph{Dining Philosophers} example (known as Philosopher in
the MCC).

\section{Architecture of MCC}
\label{sec:architecture-mcc}
The \mcc\ tool is a standalone Go program built from three main
software components\footnote{See the documentation
  at~\url{https://godoc.org/github.com/dalzilio/mcc}.} (called
\emph{packages} in Go): \vv{pnml}, \vv{hlnet}, and \vv{corenet}.
Basically, the architecture of \mcc\ is designed to resemble that of a
compiler that translates high-level code (HLPN) into low-level
instructions (P/T net). We follow a traditional structure with three
stages where: \vv{pnml} corresponds to the front-end (responsible for
syntax and semantics analysis); \vv{hlnet} provides the intermediate
representation; and \vv{corenet} is the back-end, which includes
functions for unfolding an \vv{hlnet} and for ``code generation''. A
last package, \vv{cmd}, contains boilerplate code for parsing
command-line parameters and manage inputs/outputs.

Each of these packages is interesting taken separately and can be
reused in other applications. Package \vv{pnml}, for instance,
includes all the types and functions necessary for parsing a PNML
file: it defines a \vv{pnml.Decoder}, which encapsulates an efficient,
UTF-8 compatible XML parser and can provide meaningful error messages
in case of problems. The \vv{hlnet} package, for its part, defines the
equivalent of an Abstract Syntax Tree data structure for PNML files.
Both of these packages can be easily reused in programs that need to
consume PNML data. In particular, they can help build a standalone
PNML parser with good error handling.

Finally, package \vv{corenet} contains the code for unfolding an
\vv{hlnet.Net} value into a \vv{corenet.Net}, which is a simple,
graph-like data structure representing a P/T net. The package also
contains the functions for marshalling a core net structure into other
formats; see function \vv{corenet.LolaWrite} for an example.  More
than compliance with the standard, \mcc\ takes care of many of the
``idiosyncrasies'' in the way PNML model are written in the MCC. For
instance we consider the case where \vv{numberof} does not declare a
multiplicity.

A tool developer that would like to adopt \mcc\ to generate a ``core
net'', using his own format, only needs to provide a similar
\vv{Write} function. In the case of the \vv{pnml} subcommand, that was
added on the last release of the tool, one hundred line of codes were
enough to add the ability to generate PNML files. A figure that is
similar to what we observed with the \vv{lola} subcommand.

\subsubsection*{Using package hlnet for drawing Colored nets.}
\begin{figure}
  \includegraphics[width=\textwidth]{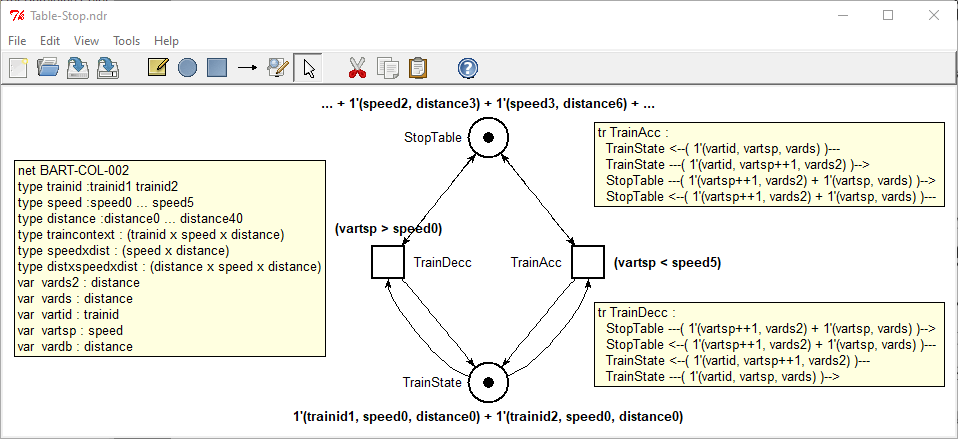}
  \caption{Result of option debug on model TrainTable, displayed in
    \vv{nd}}
  \label{fig:4}
\end{figure}

Package \vv{hlnet} also includes a function to output a textual
representation of an AST that is compatible with TINA's net syntax. It
generates a net that includes all the places and transitions in a
colored model as if it was a P/T net and uses labels to display the
expressions associated with transitions and the initial marking of
places. The net also includes ``nodes'' (comments similar to sticky
notes) for information about types, variables and arc
inscriptions. The result can be displayed and modified with \vv{nd},
the \emph{NetDraw} graphical editor distributed with TINA. We show
such an example in the screen capture of Fig.~\ref{fig:4}, which is
obtained by using \mcc\ with option \vv{-{}-debug} on the HLPN model
\emph{TrainTable} of Fig.~\ref{fig:2}.

While modifications cannot be saved back into PNML, this capability is
still useful to inspect colored model (and is often more accurate than
the graphical information included in the ``cover flow'' provided with
every model). We can also use the export function included in \vv{nd}
to generate a \LaTeX (\emph{tikz}) representation of the net. This is
what we used to generate an initial version of the diagrams that
appear in Fig.~\ref{fig:1} to~\ref{fig:3} of this paper.

\section{Comparison with other Tools}
\label{sec:comp-with-other}

The problem of (efficiently) unfolding colored models has been
abundantly covered in the literature and many of the proposed
algorithms have been implemented. We can cite the works of
M{\"a}kel{\"a}, with his tool
MARIA~\cite{makela2001optimising}; of Heiner et
al. with Marcie~\cite{liu2012efficient,heiner2013marcie}; or the work
of Kordon et al.~\cite{kordon2006optimized}, that makes a clever use
of decision diagrams in order to compute results for very large
instances. This approach is implemented in CPN-AMI~\cite{hamez2006new}
and provides the reference for P/T instances derived from Colored
model in the MCC.  All these works provide good motivations for why it
may be useful to unfold a HLPN instead of trying to analyze it
directly.

We decided to compare \mcc\ with three tools that participated in the
Model-Checking Contest: Tapaal~\cite{david2012tapaal} (with its
\vv{verifypn} tool); Marcie~\cite{heiner2013marcie} (with the
\vv{andl\_converter}); and GreatSPN~\cite{amparore201630} (that
includes a Java based unfolding tool in its editor). Since each tool
is tailored for a different toolchain---and therefore generate very
different results---it is difficult to make a precise comparison of
the performances. Hence these results should only be interpreted as a
rough estimate. For instance, Tapaal is the only tool in this list
that do not output the unfolded net on disk. This means that its
computation time do not include the time spent marshalling the result
and printing it on file.
\newcommand{\minpgfutilensuremath}[1]{\textcolor{blue}{\pgfutilensuremath{\mathbf{#1}}}}
\newcommand{\TIMEOUT}{\colorbox{red!10!white}{---}}
{\setlength{\tabcolsep}{4pt} 
  \begin{table}[t]
    \centering
    \begin {tabular}{|l||r|r||r|r|r|r|}%

      \hline\textsc {Model}&\textsc {Places}&\textsc {Trans.}&\textsc {MCC}&\textsc {Tapaal}&\textsc {Marcie}&\textsc {GSPN}\\\hline\hline %
      GlobalResAllocation-07&\pgfutilensuremath {133}&\pgfutilensuremath {291\,067}&\minpgfutilensuremath {1.7}&\pgfutilensuremath {3}&\pgfutilensuremath {14.4}&\pgfutilensuremath {22.3}\\%
      GlobalResAllocation-11&\pgfutilensuremath {297}&\pgfutilensuremath {2.10^6}&\minpgfutilensuremath {15.1}&\pgfutilensuremath {29.3}&\pgfutilensuremath {144.6}&\TIMEOUT\\%

      \hline {}DrinkVendingMachine-16&\pgfutilensuremath {192}&\pgfutilensuremath {10^6}&\pgfutilensuremath {15.5}&\pgfutilensuremath {10.7}&\pgfutilensuremath {52.8}&\pgfutilensuremath {108.1}\\%
      DrinkVendingMachine-24&\pgfutilensuremath {288}&\pgfutilensuremath {8.10^6}&\pgfutilensuremath {97.1}&\pgfutilensuremath {95.9}&\TIMEOUT&\TIMEOUT\\%
      \hline PhilosophersDyn-50&\pgfutilensuremath {2\,850}&\pgfutilensuremath {255\,150}&\minpgfutilensuremath {1}&\pgfutilensuremath {2.1}&\pgfutilensuremath {11.1}&\pgfutilensuremath {15.7}\\%
      PhilosophersDyn-80&\pgfutilensuremath {6\,960}&\pgfutilensuremath {10^6}&\minpgfutilensuremath {4.1}&\pgfutilensuremath {9.9}&\pgfutilensuremath {55.9}&\pgfutilensuremath {61.0}\\%

      \hline{}Diffusion-D050&\pgfutilensuremath {2\,500}&\pgfutilensuremath {8\,109}&\pgfutilensuremath {14.5}&\minpgfutilensuremath {0.6}&\pgfutilensuremath {4.1}&\TIMEOUT\\%
      Diffusion-D100&\pgfutilensuremath {10\,000}&\pgfutilensuremath {31\,209}&\pgfutilensuremath {243.3}&\minpgfutilensuremath {8.6}&\pgfutilensuremath {31.3}&\TIMEOUT\\%
      
      \hline{}TokenRing-100&\pgfutilensuremath {10\,201}&\pgfutilensuremath {10^6}&\minpgfutilensuremath {4}&\pgfutilensuremath {8.2}&\pgfutilensuremath {33.5}&\pgfutilensuremath {49.3}\\%
      TokenRing-200&\pgfutilensuremath {40\,401}&\pgfutilensuremath {8.10^6}&\minpgfutilensuremath {67.4}&\pgfutilensuremath {166.1}&\TIMEOUT&\TIMEOUT\\%
      \hline {}SafeBus-50&\pgfutilensuremath {5\,606}&\pgfutilensuremath {140\,251}&\pgfutilensuremath {14.2}&\minpgfutilensuremath {1.4}&\pgfutilensuremath {6.2}&\pgfutilensuremath {25.1}\\%
      SafeBus-80&\pgfutilensuremath {13\,766}&\pgfutilensuremath {550\,801}&\pgfutilensuremath {89.5}&\minpgfutilensuremath {7}&\pgfutilensuremath {20.6}&\pgfutilensuremath {133.1}\\%

      \hline {}TrainTable-Dist&\pgfutilensuremath {722}&\pgfutilensuremath {602}&\minpgfutilensuremath {1.4}&\pgfutilensuremath
                                                                                                              {12.6}&\pgfutilensuremath
                                                                                                                      {59.5}&\pgfutilensuremath
                                                                                                                              {69.4}\\%
      {}TrainTable-Stop+Dist&\pgfutilensuremath
                              {728}&\pgfutilensuremath
                                     {602}&\minpgfutilensuremath
                                            {2.1}&\TIMEOUT&\TIMEOUT&\TIMEOUT\\%
      BART-002&\pgfutilensuremath {764}&\pgfutilensuremath {646}&\minpgfutilensuremath {3.1}&\TIMEOUT&\TIMEOUT&\TIMEOUT\\%
      BART-060&\pgfutilensuremath {15\,032}&\pgfutilensuremath {19\,380}&\minpgfutilensuremath {3.2}&\TIMEOUT&\TIMEOUT&\TIMEOUT\\ %

      \hline {}SharedMemory-000200&\pgfutilensuremath {40\,801}&\pgfutilensuremath {80\,400}&\minpgfutilensuremath {0.3}&\pgfutilensuremath {1.7}&\pgfutilensuremath {2.6}&\pgfutilensuremath {5.1}\\%
      SharedMemory-001000&\pgfutilensuremath {10^6}&\pgfutilensuremath {2.10^6}&\minpgfutilensuremath {8.9}&\TIMEOUT&\pgfutilensuremath {60.3}&\pgfutilensuremath {160.2}\\%
      SharedMemory-002000&\pgfutilensuremath
                           {4.10^6}&\pgfutilensuremath
                                          {8.10^6}&\minpgfutilensuremath
                                                         {55.3}&\TIMEOUT&\TIMEOUT&\TIMEOUT\\%

      \hline {}FamilyReunion-L800&\pgfutilensuremath {2.10^6}&\pgfutilensuremath {2.10^6}&\minpgfutilensuremath {5.5}&\TIMEOUT&\pgfutilensuremath {84.8}&\pgfutilensuremath {143.0}\\%
      FamilyReunion-L3000&\pgfutilensuremath {28.10^6}&\pgfutilensuremath {27.10^6}&\minpgfutilensuremath {89.5}&\TIMEOUT&\TIMEOUT&\TIMEOUT\\\hline %

      \multicolumn{7}{c}{ }\\[-0.5em]      

      \hline {}Swap-P010000&\pgfutilensuremath
                             {20\,000}&\pgfutilensuremath
                                        {20\,000}&\pgfutilensuremath
                                                   {0.1}/\pgfutilensuremath
                                                   {0.6} &\pgfutilensuremath
                                                                {0.4}&\pgfutilensuremath
                                                                       {0.9}&\pgfutilensuremath
                                                                              {5.0}\\%
      
      Swap-P100000&\pgfutilensuremath {200\,000}&\pgfutilensuremath
                                                  {200\,000}&\pgfutilensuremath
                                                              {0.4}/\pgfutilensuremath
                                                              {4.8}&\pgfutilensuremath {26.1}&\pgfutilensuremath {15.7}&\TIMEOUT\\\hline%

      \end {tabular}\\[1em]
      \caption{Execution time (in \si{\second}) when unfolding complex PNML instances\label{tab:1}}
    \end{table}
}

\subsubsection*{Unfolding algorithm.}
We follow a very basic strategy. For each place $p$, of type say $T$,
we create one instance of $p$ for any value that inhabits $T$. (This
part is common to most of the existing unfolding algorithms.) For each
transition, $t$, we consider the set of variables occurring in the
inscription of arcs attached to it (its \emph{environment}). Then we
enumerate all possible valuations of the environment and keep only
those that satisfy the conditions associated with $t$.

Our main optimization is to follow a ``constraint solving'' approach
where we can avoid enumerating a large part of the possible
assignments when we know that the condition cannot be satisfied.  For
instance a subexpression in a conjunction is falsified. This is a less
sophisticated approach than those described in existing
works~\cite{liu2012efficient,makela2001optimising}. For instance, we
do not try to detect particular kind of expressions where an
unification-based approach could have better performances.

Typically, we should perform badly with instances similar to
Diffusion, where we may fail to cut down the size of our search
space. On the other hand, our approach is not hindered when we need to
deal with complex expressions, such as with TrainTable, that involve
at the same time tuples, \vv{successor}, and \vv{add}. Actually, our
approach may also work with nonlinear patterns, where the same
variable is reused in the same expression. Finally, all the algorithms
should work equally well on examples like Swap, because of its
simplicity.

Even if our approach is quite rustic, our experiments shows that this
does not hinder our performances. This may be because few of the
colored instances in the MCC fall in the category where clever
algorithms shine the most.

\subsubsection*{Benchmarks.}
We selected instances, with a processing time of over a second, from
different models listed in the MCC repository~\cite{HillahK17} and
from the three examples in Sect.~\ref{sec:architecture-mcc}. We give
the results of our experiment in Table~\ref{tab:1}. Computations were
performed with a time limit of \SI{5}{\minute} and a limit of
\SI{16}{\giga\byte} of RAM. In each case we give the number of places
and transitions in the unfolded net and highlight the best time (when
there is a significant difference). An absence of values ($\TIMEOUT$)
means a timeout.

Tapaal shows very good performances on many instances and
significantly outperforms \mcc\ for models {SafeBus} and
{Diffusion}. On the opposite, we see that many instances can only be
processed with \mcc. This is the case with model {BART} (even with a
time limit of \SI{1}{\hour}). Other interesting examples are models
{SharedMemory} and {FamilyReunion}. This suggest that we could further
improve our tool by including some of the optimizations used in
\vv{verifytpn} that seems to be orthogonal to what we have implemented
so far. We describe two of the optimizations performed by \mcc\ below.

Actually, sheer performance is not our main goal. We rather seek to
return a result for all the colored instances used in the MCC in a
sensible time. (Who needs to unfold a model too big to be analyzed
anyways?) At present, there are $193$ \emph{instances} of Colored nets
in the MCC repository, organized into $23$ different classes, simply
referred to as \emph{models}. We can return a result for $184$ of
these instances, with the condition of the competition. Moreover, to
the best of our knowledge, \mcc\ is the only tool able to return
results (for at least one instance) in all the models. But some
instances, like {DrinkVendingMachine-48}, should stay out of reach for
a long time, mostly due to memory space limitations.

\subsubsection*{Use of colored invariants.} The first ``additional''
optimization added to \mcc\ explains our good result on models such as
BART. The idea is to identify invariant places; meaning places whose
marking cannot be changed by firing a transition. A sufficient
condition for place $p$ to be invariant is if, for every transition
$t$, there is an arc with inscription $e$ from $p$ to $t$ iff there is
an arc from $t$ to $p$ with inscription $e'$ equivalent to
$e$. (Syntactical equality between $e$ and $e' $ is enough for our
purpose.) We say that such places are \emph{stable}; a concept
equivalent to ``test arcs'' for an HLPN. This is the case, for
example, for place {StopTable} in model {TrainTable}. When a place is
stable, we know that its marking is fixed. This can significantly
reduce the set of assignments that need
to be enumerated.\\

\subsubsection*{Use of a Petri scripting language.} The effect of our
second improvement can be observed in model {Swap}
(Fig.~\ref{fig:3}). In this case, like with model Philosopher of the
MCC, it is possible to detect that the unfolded net is the composition
of $n$ copies of the same component; where $n =
|\mathrm{Resource}|$. Each component $x$ (with
$x \in \mathrm{Resource}$) is a net with a local copy of the
places. As for the transitions, we need to keep one copy for each
``local interactions'' (such as $t_2$) and two copies for distant
interactions ($t_1$): one for the pair of components $(x\omm, x)$; the
other for the pair $(x, x\opp)$. Since type Resource is a cyclic
enumeration---this is basically a ``scalar set''---the composition of
all these components form a ring architecture.

Our tool is able to recognize this situation automatically. In such a
case we output a result that uses the \emph{TPN format}, a scripting
language for Petri net supported by the TINA toolchain. This scripting
language includes operators for make copies of net; add and rename
places and transitions; compute the product or chaining of nets;
\dots\ It also provides higher-order composition patterns, such as
pools or rings of components. We use the latter for model Swap.

Our benchmarks of Table~\ref{tab:1} include the results on two
instances of model Swap. The computation time for \mcc\ (the first
value) is mostly independent from the size of the instance (the only
difference is in parsing the PNML file.) This result conceals a much
more complex realty. Indeed, a tool that consumes a TPN script still
needs to ``expand it''. This is why we added a second value in
Table~\ref{tab:1}, which is the time taken to generate the result
using the \vv{mcc pnml} command. For information, the size of the PNML
result for model {Swap-P100000} is \SI{99}{\mega\byte}, while it is
only {200} bytes for the TPN version.

\section{Conclusion}
\label{sec:conclusion}

Tool \mcc\ is a new solution to an old problem. It is also an
unassuming tool, that focuses on a single, very narrow
task. Nonetheless, we believe that it can still be of interest for the
Petri net community, and beyond, by enriching the PNML ecosystem. As a
matter of fact, there has been a total of $26$ verification tools to
participate to the MCC since its beginning~\cite{mcc2019}, not all
``Petri tools''. Many of these tools could benefit from using \mcc.

Development on \mcc\ started in 2017, as a pet project for studying
the suitability of the Go programming language to develop formal
verification tools. Our assessment in this regard is very positive:
performances are competitive with regards to
{{C\nolinebreak[4]\hspace{-.05em}\raisebox{.4ex}{\tiny\bf ++}}}, with
good code productivity and mature software libraries; building
executables for multiple platforms and distributing code is easy;
\dots\

Since then, work has progressed steadily in-between each edition of
the MCC, with a focus on stability of the tool and on compliance with
the PNML standard. Three iterations later, \mcc\ is now sufficiently
mature to gain more exposure and provides a good showcase for an
efficient PNML parser written in Go. But \mcc\ is more than
that. First, \mcc\ was designed to lower the work needed by developers
wanting to engage in the Model-Checking Contest. It also provides new
features, such as the ability to display an interactive (read-only),
graphical view of a PNML model; see Fig.~\ref{fig:4}. Finally, it
provides a testbed for evaluating new unfolding algorithms (we show
two of these ideas in Sect.~\ref{sec:comp-with-other}).

In the future, we plan to enrich \mcc\ by computing interesting
properties of the models during unfolding. For example by computing
invariants or by finding sets of places that can be clustered
together. In that respect, the possibility to identify HLPN that can
be expressed using a ``Petri net scripting language'' could
potentially leads to new advances. For example to simplify the
detection of symmetries, something that we have been working on
recently in the context of Time Petri nets~\cite{scp2016symmetries}.

\bibliographystyle{splncs04}
\bibliography{biblio}
\end{document}